\documentclass[11pt]{article}

\usepackage{graphicx}
\usepackage{dcolumn}
\usepackage{amssymb}
\usepackage{bm}
\usepackage{amsfonts}
\usepackage{amsmath}
\usepackage{bbm}
\usepackage[T1]{fontenc}
\usepackage[latin1]{inputenc}

\date{\today}

\begin{document}

\title{Omnidirectional Gravitational Wave  Detector\\with  a Laser-Interferometric Gravitational Compass}

\author{ M.D. Maia\thanks{maia@unb.br}  and Ivan S. Ferreira\thanks{Ivan@fis.unb.br}\\{\small University of Brasilia,  Institute of Physics, Brasilia 70910-900}\\
 Claudio M. G. Sousa\thanks{desousa@ucb.br},\\{\small  Catholic University, Brasilia, 71966-900 \& Federal University of Para, Santarem,  68040-070}\\ 
Nadja S. Magalh\~aes\thanks{nadjasm@gmail.com},\\ {\small S\~ao Paulo Federal University, DCET, Diadema, 09913-030}\\
Carlos Frajuca\thanks{frajuca@gmail.com},\\
{\small Instituto Nacional de Pesquisas Espaciais, S\~ao Jos\'e dos Campos,  S.P. 12227-010}}

\maketitle

\begin{abstract}
Based on  the  Szekeres-Pirani gravitational compass we  suggest  the addition of  a fourth, non-coplanar   mass/mirror  to the presently existing  laser   based gravitational wave observatories,  enabling them to operate omnidirectionally, to  filter out  ambiguous interpretations and  to   point out  the  direction of the gravitational wave source.  
\end{abstract}

\section{Introduction}
The recent  detection  of gravitational waves  in the event GW150914,  clearly shows the  expected waveforms  produced by  a coalescent  black hole  binary system \cite{Abbot}. The  sophisticated  upgrade  implemented in  those detectors proved to  be   adequate to produce the very  small  shifts of the  interferometer fringes,  of the order of two hundredth of the proton diameter.

 The theory  behind the  gravitational wave observatories,  is   based on the well known geodesic deviation equation \cite{Pirani1967}.  Although  this equation by itself  is  a purely  geometrical statement in Riemannian  geometry, its  application to the space-times of General Relativity provides an intuitive view on how the gravitational field  affects the  displacement of test particles in terms  of the  variations of the curvature of  the  space-time in their  immediate  neighborhood.  To see how this works, consider initially  two  massive  test particles  at positions  A and B,  subjected  to the Earth's gravitational field. The particle  at A sends a signal with velocity $P$ to the particle at B along the shortest possible distance, meaning that  the signal travels along a geodesic with equation  $\nabla_P P = 0$. Immediately  after a  signal is  sent,  both particles  are allowed to fall freely (that is, under  the exclusive influence of the  gravitational  field),  along  time-like geodesics with tangent vectors  $T$ and  $T'$  respectively, with $T'$  parallel to $T$,   also satisfying   geodesic equations $\nabla_T T = 0$ and $\nabla_{T'} T'=0$. After a  while,  the particles reach positions A' and B', where   B'  is reached by the signal  emitted from the particle at  A'. Since the  signal  and  fall  speeds are  different, the first   needs to  vary along the fall,  at a rate  such that at the end  we obtain a closed  geodesic parallelogram  satisfying the closing condition $\nabla_T P=\nabla_P T$.  In  operator form, the Riemann tensor $R$ evaluated in such parallelogram  gives 
\begin{equation} 
R(T,P)T = [\nabla_T ,\nabla_P ]T = \nabla_T (\nabla_P T) -\nabla_P (\nabla_T T) = \nabla_T (\nabla_T P).
\end{equation}
Denoting by $a = \nabla_T P$ the acceleration of the signal along the free fall, this  gives   the geodesic deviation equation (GDE)
\begin{equation}
\nabla_T\, a = R(T,P)T.  \label{eq:GDE}
\end{equation} 
A  more familiar form of this equation follows  from its application to  a reference frame  defined by base vectors  $\{e_\mu\}$  with $\mu = 0,1,2,3 $, in which the components of the Riemann tensor are given by $R(e_\mu,e_\nu)e_{\rho} = R_{\mu\nu\rho\sigma}e^\sigma$. In particular,  taking $T = e_0$ as the time direction,  and $P = e_i$, where  $e_i$,  is any  one  of the  space-like base vectors,  (\ref{eq:GDE})  becomes
\begin{equation}
\frac{d a^i}{cdt}= R_{0i0i}, \;\;\;   \mbox{ fixed i}, \label{eq:GDEcomponents}
\end{equation}
where  $c$ is the speed of light. Thus, the  falling pair of particles  describe  a   2-dimensional  world-sheet  in space-time  with  signature $(-,+)$,  whose  curvature tensor  has  a  single component  $R_{0i0i}$,  producing the  variation of the signal acceleration  in the left hand  side.

For gravitational wave detection we use the same reasoning,  where again  the  two   particles are initially at rest, either floating in the void, or properly  suspended to compensate  for the Earth's  gravitational pull.  When  they are hit  by a gravitational wave  train,  instead of  experiencing a  ``free  fall",  they  will  ``surf the gravitational waves" along   geodesics of  the locally distorted space-time  geometry.  As the  geodesic parallelogram closes,  we  obtain the same   equation (\ref{eq:GDE}),  defined in the  world-sheet generated  by  the  geodesics with tangent vectors   $T$ and $P$. Again,  using  a reference frame  with  $T=e_0$  and  $P=e_i$, for any space-like vector  $e_i$,   we obtain the same    equation  (\ref{eq:GDEcomponents})  describing the  geodesic  deviation in the the 2-dimensional world-sheet generated by  $\{e_0,e_i\}$.        

As it happens, any  2-dimensional Riemannian (or pseudo-Riemannian)  manifold  is  conformally flat. This means that   it is  always  possible to find  a  function $\varphi$ such that  its  metric  can be written as $g_{ij}=e^{2\varphi} \eta_{ij}$  and  $g^{ij}=e^{-2\varphi} \eta_{ij}$,  where $\eta_{ij}$  denotes the  2-dimensional  metric for  $i,j =1,2$. Calculating  the  Ricci tensor  for   $g_{ij}$  we obtain $R_{ij} = (\varphi_{,11}-\varphi_{,22})\eta_{ij}$  and  the Ricci  scalar gives   $R=2e^{-2\varphi}(\varphi_{,11}-\varphi_{,22})$.  It follows that $R_{ij} \equiv \frac{1}{2}R g_{ij}$,  so that    Einstein's  equations vanish  identically in any 2-dimensional space-time.  Consequently,  the interpretations of a  gravitational wave event in detectors  constructed with only two test masses may  require  considerations on   2-dimensional gravitational  theories  which are not derived from  Einstein's  equations,  like   in the  Jackiw-Teitelboim, in the Liouvile gravity  and  in  string-like theories \cite{Maia}.  
 
The presently existing interferometric  gravitational wave  detectors  based on three,  non-coplanar, test  masses (e. g. like   GEO600, VIRGO and LIGO) represent an improvement of  the  sensibility of the  2-mass  systems obtained by  the addition of one  extra  test mass,  or  equivalently,  by  increasing the  dimension  of the previous  world-sheet   to a 3-dimensional  world-volume  with signature  $(-,+,+)$ within the four-dimensional  space-time, in which  the  geodesic deviation  takes place.  Actually, the  LIGO  equipments  use  4  masses,  with   two  at the extremes  of each arm  $L_1$  and  $L_2$  and  two extra  masses located near  the  intersection of those  arms,  all  positioned  in the same plane, so that for  point of  view  of  the geodesic deviation equation   effectively takes place  in a 3-dimensional space-time. 

 After being hit  by a gravitational wave front, the deformation of the  space-time geometry is  given by the  components of the Riemann tensor of the  3-dimensional space-time  volume  generated by the  motion of  the  3  masses. The geodesic deviation equation   describe the  variation of the   communication  signals between  the  three   masses as compared with the initial positions. Two geodesic signals  appear, with  accelerations  denoted by $a_{ij}$, with $i,j = 1,2$. This may also be  expressed in terms of the frequency shifts, $e^{i\omega_{ij}}$,  so that  (\ref{eq:GDEcomponents}) reads"
\begin{equation}
\frac{1}{\omega_{ij}}\frac{d \omega^{ij}}{cdt}= R_{0i0j},\;\; i =1,2.  \label{eq:omega}
\end{equation}

Although  there is  an improvement with  respect to the  2-mass detectors,  the  restriction of General Relativity to an  effective  3-dimensional space-time subset implies  that the  curvature changes  takes occurs with  a smaller number of  degrees of freedom as  compared  to the complete  four dimensional General Relativity.  This is  a  consequence of  the fact that  the Riemann tensor in 3-dimensions is determined  by  the Ricci tensor \cite{Carlip1}. Therefore, when Einstein's  equations  are  applied,  the  Riemann  curvature  becomes  essentially determined by matter, so that in the case of  empty space, the  Riemann tensor vanishes.  However, it is possible to  restore  a  curvature  term  in  empty space,  by the use of a non-trivial (Chern-Simmons)  topological construction,  in association with  topological   BTZ  black-holes 
\cite{Banados,Carlip2}. Since the LIGO  detector  falls in the category of  3-mass  system,  the  event GW140915   may correspond to a  genuine gravitational wave  from four-dimensional Einstein's relativity,  but it does not exclude the possibility that a 2+1  dimensional  topological  gravitational  theory  may  have occured.  However, this is  something still  to be  checked.

\section{The Gravitational Compass}

The  observables of Einstein's gravitational field are  given by the  eigenvalues of the Riemann  tensor
\begin{equation}
R_{\mu\nu\rho\sigma}X^{\mu\nu}= \lambda X_{\rho\sigma}, \label{eq:eigenvalues} 
\end{equation}
where   $X^{\mu\nu}= X^{[\mu\nu]}$  is a  two-form  (or a bi-vector   or  a  skew-vector) eigenvector \cite{Pirani1957,Pirani1962}.  In a four-dimensional space-time,  there are at most  six  independent  eigenvectors, solutions of (\ref{eq:eigenvalues}). 
These  eigenvectors  can  be   written in the Newmann-Penrose null  frame  to determine  the  six  Petrov types  O, I, II, III, N and D of the the gravitational field.  Type O  corresponds to zero eigenvalues, so that    we  have  only  five  effective Petrov types corresponding  to the  five  non-trivial  observable degrees of freedom of the  gravitational field.  

On the other hand, the  eigenvectors of  the curvature tensor  generate  a  six  dimensional  space,  which can be set  in a  1:1 correspondence with six linearly independent 3 $\times 3$   matrices.   Using such matrices,  it is possible to  determine the   principal curvature directions of the gravitational field  \cite{Lenzi}. This is  the basis  of   Peter Szekeres' proposition  of a  device  called \textit{the gravitational compass}, consisting of  four non-coplanar masses, labeled O, A, B, C, connected by six dynamometers whose forces measure the eigenvalues of the gravitational field (fig. \ref{fig:compass} below). The compass could in principle be  rotated around the mass at O, so that the  forces on the  diagonal dynamometers  AB, AC, BC   vanish,  and   the  radial dynamometers OA, OB, OC  indicate the principal  directions of the curvature tensor \cite{Szekeres}. 
The gravitational compass  offers  a  practical  measurement of the true  relativistic gravitational field,  by  simply placing   such  device in a region where the presence of a gravitational field is  measured through the forces given by the six dynamometers.
\begin{figure}[!h]
\begin{center}
\includegraphics[width=5cm,height=5cm]{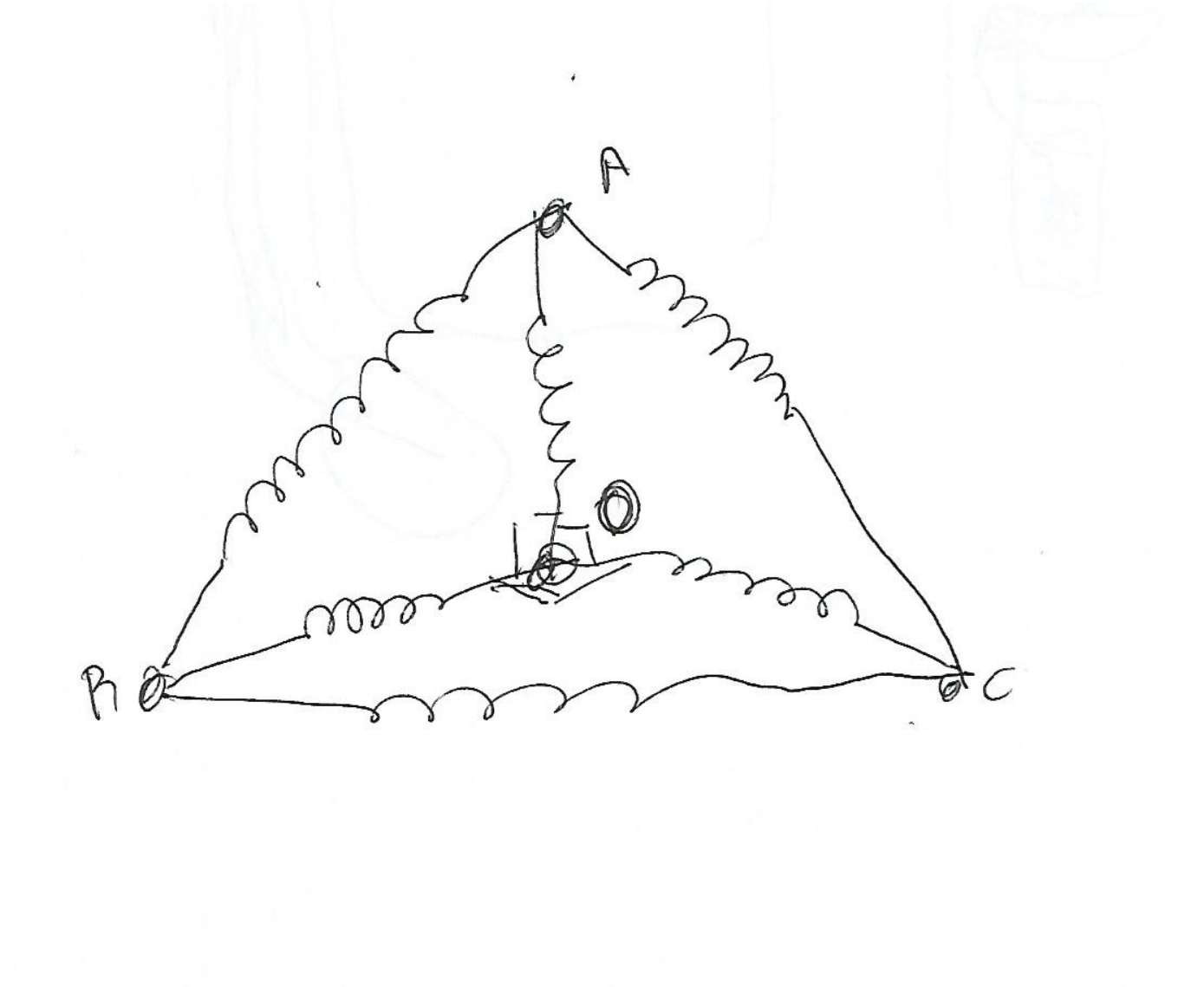}
\caption{Szekeres' Gravitational Compass}
\label{fig:compass}
\end{center}
\end{figure}

 The  resonant-mass spherical gravitational wave detector  constructed with  with a solid  sphere  made of an elastic material can be thought as  a  practical realization of  Szekeres'  compass,  where the dynamometers are  replaced by the 
tensions resulting from of the  principal modes of oscillation of  the  spherical surface  \cite{Johnson,Aguiar}.

A simpler  device of the same  nature of the Szekeres compass, was proposed by  Pirani,   in which the  dynamometers are  replaced by  a  telemetric measure   of the relative displacements of the four points of the Szekeres gravitational compass  \cite{Pirani1967}. Since  four  points  in  the  3-dimensional space determine  a  spherical surface, the  Szekeres-Pirani   gravitational compass could  provide a practical  measure of the   gravitational  field by  observing the principal  modes  of oscillation  of the four masses, or  equivalently of the  sphere,  using laser  interferometers.  Such modes are illustrated  below  by  the  graphical  display of  the p-modes of a  neutron star  oscillation \cite{Nilson}. 
\begin{figure}[!h]
\begin{center}
\includegraphics[width=6cm,height=5cm]{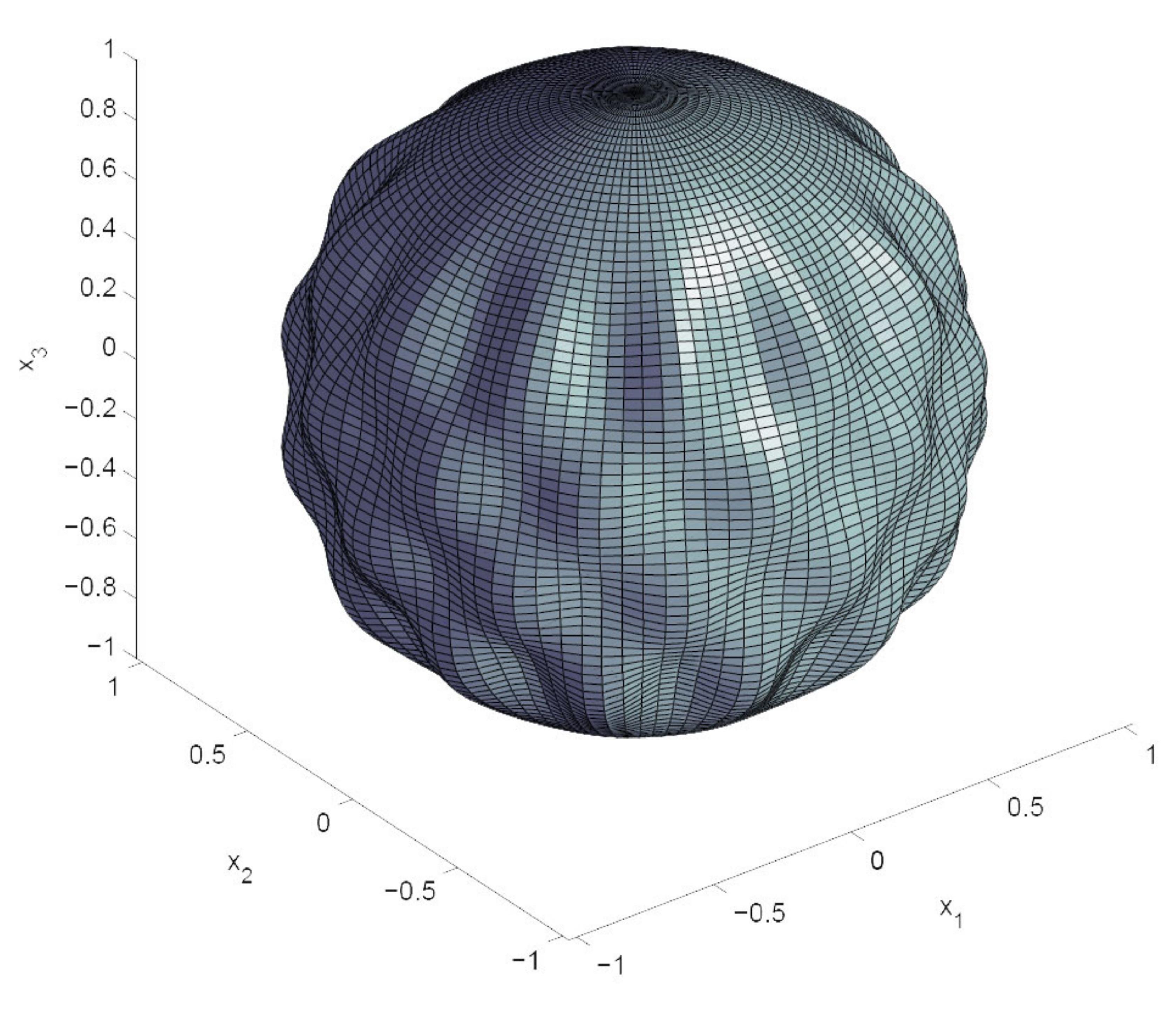}
\caption{Oscillations in a spherical surface.} 
\end{center}
\end{figure}
The   gravitational compass  owes its  name  to its  ability  to  find  the   direction of a  given   source of gravitation,   by determining  the  principal  directions of the  curvature tensor. Using the symmetry  group of the above mentioned matrices, we can find that  when  the   forces along the  diagonal  dynamometers AB, AC  and  BC vanish,  the  forces  along OA, OB  and  OC  correspond to the principal directions  of the curvature  tensor, making it  a  natural device to measure  gravitational radiation, pointing out to  the direction of the source.  
Thus, in principle it is  possible to construct  an omnidirectional interferometric gravitational detector by  extending  some of  presently existing  laser-interferometric   gravitational wave  detectors through   the addition of a  fourth mass  outside the   plane of  the  three existing ones \cite{Nadja}.

As an  example,  the GEO600   detector  in  Hanover, Germany,  could   be  upgraded to a gravitational compass  by  adding   a  600m tower   in the  same detector site (The construction of  such towers made of  metal mesh  supported by an  array of stays has  proven to be feasible by the  operating  650m  radio antenna  in Poland.). Of course, some   additional  engineering challenges woud be  in the way,  such  as  the inclusion of  a well insulated  tube for the laser to compensate for the temperature gradient,  the  suspension of the mirror at such altitudes,  and the swing of the towers caused by strong  winds. Alternatively, instead of  a tower a  deep  600m  well  could  be  dug,  but  this  would  probably  need  a new  construction site with  additional  costs  and   engineering adaptations.  

In principle the  Szekeres-Pirani  Gravitational Compass may also be  applied  to the  planed LISA space detectors of gravitational waves,  which   would have  the same   limitations of the 3-mass  plane geometry of the present laser  detectors.  In order to make the detector omnidirectional a  fourth sattelite would be required,   again defining a spherical-like detector in space.  Admittedly, the  stability of  a  four  body   system  in space may  represent  an additional  problem    as the  satellites may depend of  a propulsion system for the  orbit  stability at least  for   long term  use.   
\vspace{1cm}\\
Summarizing, the  observables of the gravitational  field  are given by the five  non-trivial eigenvalues of  the curvature tensor  of  space-time. Therefore  an   efficient gravitational wave detector  truly in conformity with Einstein's gravitational theory   must  be able to  detect  the  corresponding  five degrees of  freedom of Einstein's gravitational field. With basis on the Szekeres-Pirani  Gravitational compass we suggest that the existing  GEO600 
laser-interferometric  gravitational wave detector could be   converted  to  a four  test-mass  gravitational wave detector,  thus becoming the  prototype for omnidirectional interferometric detectors. As well, we suggest that planned interferometric detectors, could be designed according to the gravitational compass concept from the beginning in order to be a fully omnidirectional gravitational wave observatory by itself".

\vspace{0.6cm}
Acknowledgments\\
The  authors   acknowledge   Dr. Odylio D. Aguiar  for suggestions  on  the
construction of  spherical gravitational  wave detectors.  NSM and CF acknowledge FAPESP for its support to their research through the thematic project 2013/26258-4.

\end{document}